\documentclass{PoS}

\title{Determination of the $A_2$ amplitude of $K\to\pi\pi$ decays}

\ShortTitle{Determination of the $A_2$ amplitude of $K\to\pi\pi$ decays}

\author{
	\speaker{T. Janowski}, C. T. Sachrajda,\\
	School of Physics and Astronomy, University of Southampton, Southampton SO17 1BJ, United Kingdom\\
        E-mail: \email{tj1g11@soton.ac.uk}, \email{cts@soton.ac.uk}}
\author{
	P. A. Boyle,\\
	SUPA, School of Physics, The University of Edinburgh, Edinburgh EH9 3JZ, United Kingdom\\
	E-mail: \email{paboyle@ph.ed.ac.uk}}
\author{
	N. H. Christ, R. D. Mawhinney, H. Yin, D. Zhang\\
	Department of Physics, Columbia University, New York, NY 10025, USA\\
	E-mail: \email{nhc@phys.columbia.edu}, \email{rdm@physics.columbia.edu}, \email{yinnht@phys.columbia.edu}, \email{dz2203@columbia.edu}
}	
\author{
	N. Garron,\\
	School of Mathematics, Trinity College, Dublin 2, Ireland\\
	E-mail: \email{ngarron@maths.tcd.ie}}
\author{
	A. T. Lytle,\\
	Dept. of Theoretical Physics, Tata Institute of Fundamental Research, Mumbai 400005, India\\
	E-mail: \email{atlytle@theory.tifr.res.in}
}

\abstract{We review the status of recent calculations by the RBC-UKQCD collaboration of the complex amplitude $A_2$, corresponding to the decay of a kaon to a two pion state with total isospin 2. 
In particular, we present preliminary results from two new ensembles: $48^3 \times 96$ with $a^{-1}=1.73$\,GeV and $64^3 \times 128$ with $a^{-1}=2.3$\,GeV, both at physical kinematics.
Both ensembles were generated Iwasaki gauge action and domain wall fermion action with 2+1 flavours. 
These results, in comparison to our earlier ones on a $32^3$ DSDR lattice with $a^{-1}=1.36$\,GeV, enable us to significantly reduce the discretization errors.
The partial cancellation between the two dominant contractions contributing to Re($A_2$) has been confirmed and
we believe that this cancellation is a major contribution to the $\Delta I=1/2$ rule.}

\FullConference{31st International Symposium on Lattice Field Theory LATTICE 2013\\
		 July 29 - August 3, 2013\\
		 Mainz, Germany}

\begin{document}

\section{Introduction}

The $\Delta I = 1/2$ rule encodes the empirical observation that the ratio of decay amplitudes $A_0$ and $A_2$ for a kaon decaying to two pions with isospin 0 and 2 is
	$\frac{\mathrm{Re} A_0}{\mathrm{Re} A_2} \approx 22.5$.
This is a surprising result, because it implies that the $I=0$ decays are 500 times more likely.
Recent results by the RBC-UKQCD collaboration have shed some light on the nature of this phenomenon~\cite{Boyle:2012ys}, although the $\Delta I =1/2$ amplitude still needs to be calculated at physical kinematics to reach conclusive quantitative results. 
In \cite{Blum:2012uk} the $\Delta I=3/2$ amplitude at physical kinematics has been calculated for the first time. 
The major limitation of this calculation was the use of a single lattice spacing, making it difficult to evaluate the systematic errors related to UV cutoff effects.
In this work we repeat the evaluation of $A_2$ on two ensembles with different lattice spacings.
This allows us to extrapolate to the continuum limit and thus reduce the error introduced by the cutoff.
All the results quoted in this paper are preliminary.

\section{Theoretical background}

		$K\rightarrow \pi\pi$ decay amplitudes are defined by:
		\begin{equation}
		A_{2/0} = \langle (\pi\pi)_{I=2/0} \mid H_W \mid K^0\rangle,
		\end{equation}
		where $H_W$ is the component of the weak Hamiltonian which changes the strangeness by one unit.
		The weak Hamiltonian can be separated into short and long distance contributions by using the operator product expansion:
		\begin{equation}
		H_W = \frac{G_F}{\sqrt{2}}V_{ud}^*V_{us}\,\sum_{i}C_i (\mu) Q_i(\mu),
		\label{eq:hw}
		\end{equation}
		where $G_F$ is the Fermi's constant, $V_{us}$ and $V_{ud}$ are CKM matrix elements, $C_i$ are the Wilson coefficients, which contain information about short distance physics and $Q_i$ are the corresponding four-quark effective operators.
		
		The nonperturbative contribution to the decay amplitude comes from the matrix elements:
		\begin{equation}
		M_i^{I=2/0} \equiv \langle (\pi\pi)_{I=2/0} \mid Q^{\Delta I = (3/2)/(1/2)}_i \mid K^0 \rangle.
		\end{equation}
		There are only 3 operators which contribute to $A_2$.
		We label them according to their chiral $SU(3)_L\times SU(3)_R$ transformation properties. We have one (27,1) operator and two electroweak penguin operators labelled (8,8) and $(8,8)_{mx}$, where the index $mx$ denotes a colour mixed operator. 
		We found that the dominant contribution to Re($A_2$) comes from (27,1) operators, while the dominant contribution to Im($A_2$) in the $\overline{\mathrm{MS}}$ scheme at 3\,GeV comes from the $(8,8)_{mx}$ operator.
		Explicit expressions for each of these operators can be found in \cite{Blum:2012uk}.
		
	One of the biggest challenges in this calculation, which is also the main limitation in the $\Delta I = 1/2$ calculation, is ensuring that the pions have physical momenta.
	In centre-of-mass frame with periodic boundary conditions, the ground-state for the two-pion system will correspond to each pion being at rest. 
Naively, obtaining results at physical kinematics would not only require a larger volume for the physical momenta to be allowed (about 6\,fm), but also make it necessary to extract the signal from a two-pion excited state.
Such a calculation would require subtraction of the ground state contribution which would introduce large statistical noise to our results.  
To avoid this problem we use antiperiodic boundary conditions for the $d$ quark (and periodic for the $u$ quark) \cite{Kim:2003xt}. The allowed momenta for the $\pi^+$ meson become $p = \pm \frac{\pi}{L}, \pm \frac{3\pi}{L},\ldots$ with the $\pi^0$ remaining at rest.
	Fortunately, in the $I=2$ case only, we can use the Wigner-Eckart theorem to relate the physical $K^+\rightarrow\pi^+\pi^0$ matrix element to an unphysical one which contains two $\pi^+$ mesons in the final state which can be created with physical momenta in the ground state.
	The relation is given by:
	\begin{equation}
		\underbrace{\langle (\pi\pi)^{I=2}_{I_3=1}\mid}_{\sqrt{2}\langle \pi^+\pi^0 \mid} Q^{\Delta I = 3/2}_{\Delta I_3 = 1/2} \mid K^+ \rangle = \sqrt{\frac{3}{2}} \underbrace{\langle (\pi\pi)^{I=2}_{I_3=2}\mid}_{\langle \pi^+\pi^+ \mid} Q^{\Delta I = 3/2}_{\Delta I_3 = 3/2} \mid K^+ \rangle\:.
	\end{equation}
	The ensembles have been tuned so that antiperiodic boundary conditions in 3 directions (which induce momentum $p=\frac{\sqrt{3}\pi}{L}$) correspond to physical kinematics.

	The Wilson coefficients and effective operators apprearing in equation (\ref{eq:hw}) are both renormalization scheme and scale dependent.
	Since the operators are evaluated on a lattice and Wilson coefficients typically in the $\overline{\mathrm{MS}}$ scheme, we need a way to relate these two schemes.
	The $\overline{\mathrm{MS}}$ scheme cannot be simulated directly on a lattice, which motivates the use of an intermediate scheme. Our choice is the RI-SMOM scheme~\cite{Sturm:2009kb}.
	The renormalization condition for the four-quark operators is given by:
	\begin{equation}
		\mathrm{Tr}(\Gamma_{Ri}(\mu)P) = 1
	\end{equation}
	where $\Gamma$ is a four point Green's function and $P$ is a suitably chosen projection operator. 
	We use the same projection operators as in~\cite{Blum:2012uk}.
	
\begin{figure}
		\begin{center}
		\includegraphics[scale=0.4]{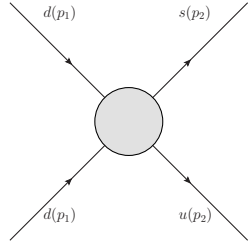}
		\caption{Momentum flow defining a renormalization condition of a four quark operator in RI-SMOM scheme. The momenta are chosen so that $p_1^2=p_2^2=(p_1-p_2)^2=\mu^2$.}
		\label{fig:rismom}
		\end{center}
	\end{figure}

	The renormalization scale is defined by a choice of momenta as shown in Figure \ref{fig:rismom}.
	The renormalized operators are related to the bare ones by:
	\begin{equation}
		Q_{Ri}(\mu) = \frac{Z_{ij}(\mu a)}{Z_q^2(\mu a)} Q_{0j}(a),
	\end{equation}
	where the indices $0$ and $R$ refer to bare (lattice) and renormalized (RI-SMOM) operators respectively and $i$ and $j$ label the operator and $Z_q$ is the quark field renormalization constant.
	Operators will, in general, mix with each other under renormalization, but chiral symmetry suppresses mixing of operators in different irreducible representations of the chiral symmetry group.
	
	Finally, we need to consider the effects of finite volume, which will result in a multiplicative correction to the matrix element \cite{Lellouch:2000pv}:
	\begin{equation}
		\langle \pi\pi \mid H_W \mid K \rangle_\infty = F \langle \pi\pi \mid H_W \mid K \rangle_{FV}\:.
	\end{equation}
	The subscripts $\infty$ and $FV$ correspond to infinite and finite volume respectively, and the factor $F$ is given by the Lellouch-L\"{u}scher formula:
	\begin{equation}
		F^2 = 4\pi \left(q\frac{\partial \phi}{\partial q} + p \frac{\partial \delta}{\partial p} \right)\frac{m_KE_{\pi\pi}}{p^3}\:,
	\end{equation}
	where $V$ is the volume, $p$ is the magnitude of the momentum of a pion in the centre of mass frame given by $p=\sqrt{\frac{E_{\pi\pi}^2}{4}-m_\pi^2}$ and $q$ is the corresponding wave number $q=pL/2\pi$. 
	$E_{\pi\pi}$ is the two-pion enegy in the $I=2$ channel and $m_\pi$ and $m_K$ are the pion and kaon masses respectively.
	The function $\phi$ is defined by the condition:
	\begin{equation}
		\tan \phi = -\frac{q \pi^{3/2} }{Z_{00}(1;q)},\quad Z_{00}(1;q) = \frac{1}{\sqrt{4\pi}} \sum_{n \in \mathbb Z^3} \frac{1}{n^2-q^2}\:,
	\end{equation}
	$\delta$ is the two-pion s-wave phase shift, which can be calculated using the L\"{u}scher quantization condition, $\delta(q) + \phi(q) = n\pi$, but the calculation of the derivative requires an approximation.
	The results presented here were obtained using the approximation that $\delta$ is linear between $0$ and $q$.
	The remaining finite volume corrections have exponential volume dependence which can be estimated using chiral perturbation theory and are treated as a systematic error as shown in section~\ref{sec:err}.

\section{Ensemble parameters}

We have generated two new 2+1 flavour ensembles: $48^3\times 96$ with $\beta = 2.13$ ($a^{-1}=1.73(1)$\,GeV) and $64^3 \times 128$ with $\beta=2.25$ ($a^{-1}=2.30(4)$\,GeV).
Both ensembles are generated with the same action, which is the Iwasaki gauge action with domain wall fermions.
The scale and quark masses were set by choosing the masses of pion, kaon and the omega baryon to be equal to their physical values.
The corresponding sea quark masses are $am_{ud} = 7.8 \times 10^{-4}$ and $am_s = 3.62\times 10^{-2}$, with the residual mass $am_{res} = 6.19(6)\times 10^{-4}$ for the $48^3$ ensemble and $am_{ud} = 6.78\times 10^{-4}$, $am_s = 2.661 \times 10^{-2}$ and $am_{res} = 2.93(8) \times 10^{-4}$ for the $64^3$ ensemble.

Note that both ensembles have the same volume with box size of about 5.49\,fm.
This ensures that the continuum extrapolation only depends on the cutoff and not on the finite volume effects.

The (preliminary) results presented here were obtained using 60 gauge configurations on the $48^3$ ensemble and 33 on the $64^3$ ensemble.
The large statistical variance one expects with a relatively small number of gauge configurations is reduced by all mode averaging \cite{Blum:2012uh} and the corresponding significant investment of effort in the analysis. 

\section{Results}
\begin{figure}
		\begin{center}
		\begin{tabular}{cc}
		\includegraphics[scale=0.5]{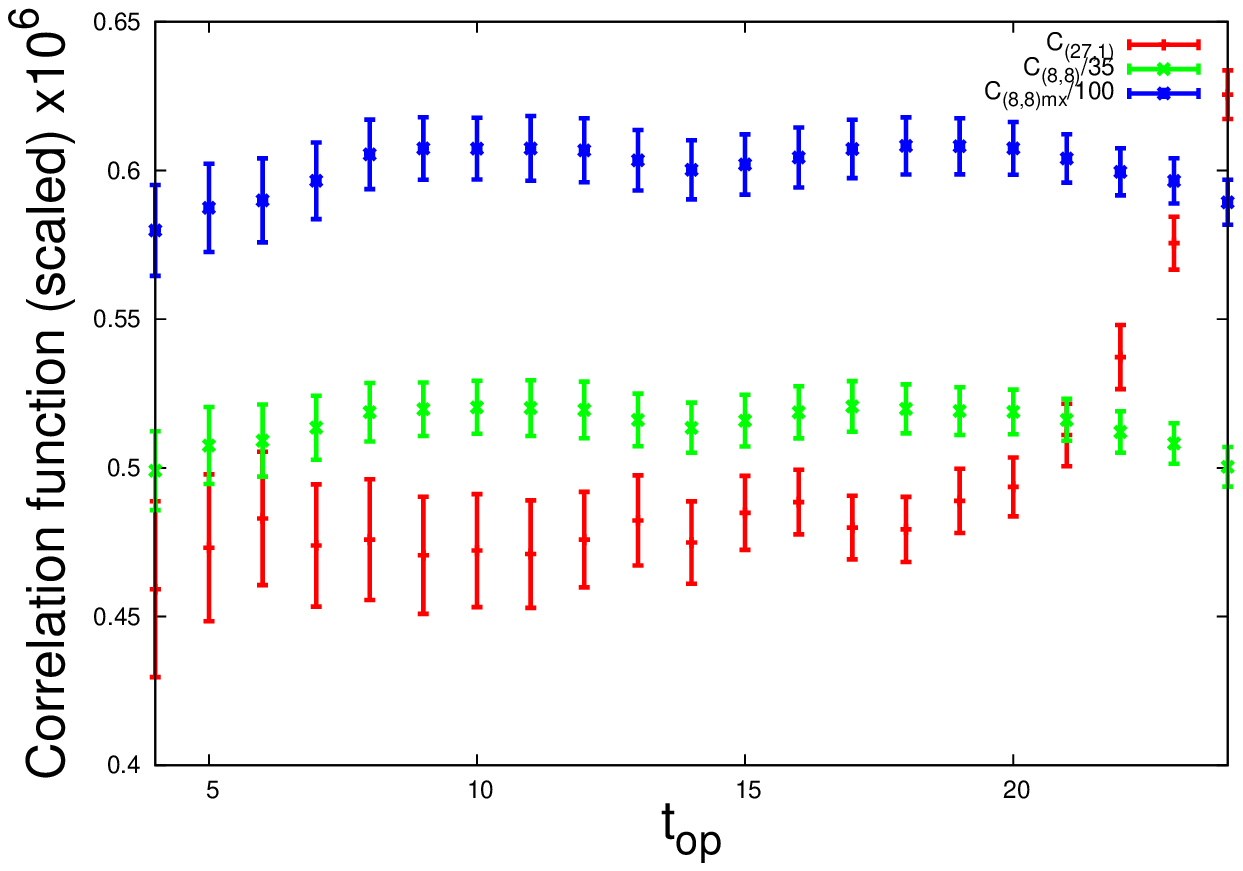} & \includegraphics[scale=0.5]{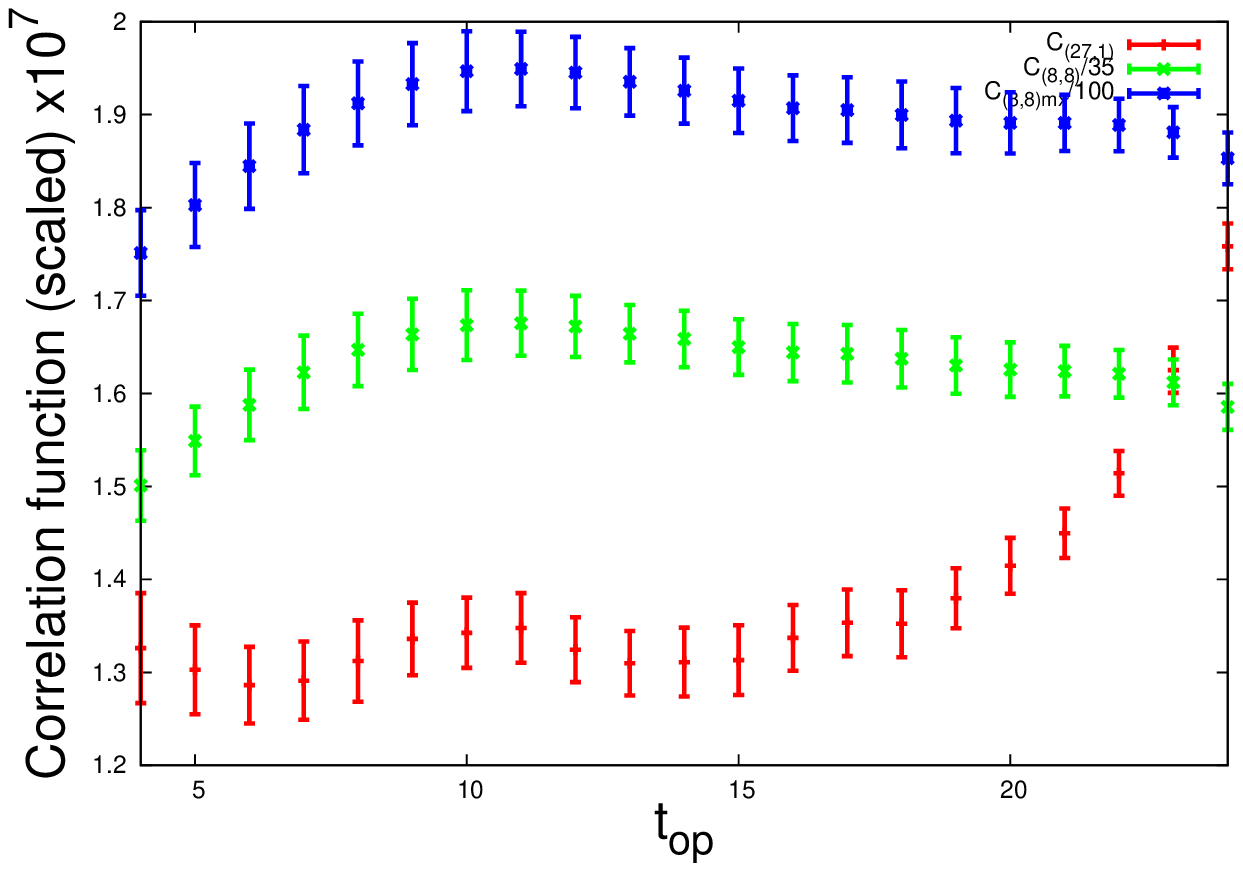} \\
		\end{tabular}
		\caption{$K\rightarrow\pi\pi$ three-point correlation function on the $48^3$ lattice (left) and $64^3$ lattice (right) with a kaon-pion separation of 26}
		\label{fig:3pt48cube}
		\end{center}
	\end{figure}

	The values of $K\rightarrow\pi\pi$ three-point correlation functions, defined for a fixed kaon-pion separation ($t_{\pi\pi}$) as:
		\begin{equation}
			C^{K\rightarrow \pi\pi}_i (t_{op}) = \langle 0 \mid \bar q(t_{\pi\pi})\gamma^5 q(t_{\pi\pi})\bar q(t_{\pi\pi})\gamma^5 q(t_{\pi\pi}) Q_i(t_{op}) \bar s(0) \gamma^5 q(0) \mid 0 \rangle
		\end{equation}
	are plotted in Figure \ref{fig:3pt48cube}, where we used $t_{\pi\pi}=26$, but an error weighed average over several values of $t_{\pi\pi}$ was used to calculate the results shown in tables \ref{tab1} and \ref{tab2}.
	The $48^3$ results are obtained by averaging over $t_{\pi\pi} \in \{ 24, 26, 28, 30, 32, 34, 36, 38\}$ and the $64^3$ results were averaged over $t_{\pi\pi} \in \{ 26, 28, 30, 32, 34, 36\}$.
	The expected time dependence of these functions is:
	\begin{equation}
		C^{K\rightarrow \pi\pi}_i(t_{op}) = N_{\pi\pi}N_{K}M_{i}e^{-(m_K-E_{\pi\pi})t_{op}}e^{-E_{\pi\pi}t_{\pi\pi}}.
	\end{equation}	
	Our ensembles have been tuned so that $E_{\pi\pi} \approx m_K$, which manifests itself by clear plateau regions in both plots.
	The corresponding bare matrix elements are shown in table \ref{tab1}.
	\begin{table}
	\begin{center}
	\begin{tabular}{|l|l|l|l|}
	\hline
	& $M_{(27,1)}^{LAT}$ & $M_{(8,8)}^{LAT}$ & $M_{(8,8)mx}^{LAT}$\\
	\hline
	$48^3$ result & $1.521(32) \times 10^{-4}$ &$5.73(12) \times 10^{-3}$ &$1.911(41) \times 10^{-2}$ \\
	$64^3$ result & $6.056(86) \times 10^{-5}$& $2.596(32) \times 10^{-3}$& $8.61(10) \times 10^{-3}$\\
	\hline
	\end{tabular}
	\end{center}
	\caption{$\Delta I = 3/2$ matrix elements in lattice regularization scheme.}
	\label{tab1}
	\end{table}
	After renormalization and correction for finite volume effects, the resulting amplitudes are shown in table \ref{tab2}.
	\begin{table}
	\begin{center}
	\begin{tabular}{|l|l|l|}
	\hline
	& $48^3$ & $64^3$ \\
	\hline
	Re($A_2$)[GeV] & $ 1.411(22) \times 10^{-8}$ & $1.398(17) \times 10^{-8}$\\
	Im($A_2$)[GeV] & $-6.40(11) \times 10^{-13}$ & $-6.438(74) \times 10^{-13}$ \\
	\hline
	\end{tabular}
	\end{center}
	\caption{Results for Re($A_2$) and Im($A_2$) before continuum extrapolation.}
	\label{tab2}
	\end{table}

\begin{figure}
		\begin{center}
		\includegraphics[scale=0.5]{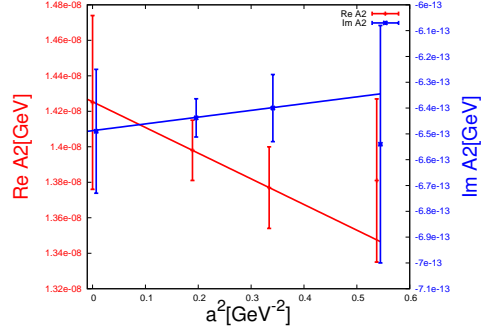}
		\caption{Taking the continuum limit for Re($A_2$) (red data points, left axis) and Im($A_2$) (blue points, right axis). The rightmost point is the $32^3$ DSDR result}
		\label{fig:contlimit}
		\end{center}
	\end{figure}

	For our choice of action we expect an $a^2$ dependence of the amplitudes on the lattice spacing.
	We can use this to obtain the continuum limit as shown in Figure \ref{fig:contlimit}, where we extrapolated through the two points corresponding to $48^3$ and $64^3$ ensembles.
Our previous result from the $32^3$ DSDR lattice was not used in the extrapolation, because the different gauge action results in a different approach of $A_2$ to the continuum limit than for the Iwasaki ensembles. The DSDR point is nevertheless, shown in Figure \ref{fig:contlimit} for completeness.

	\begin{figure}
		\begin{center}
		\includegraphics[scale=0.5]{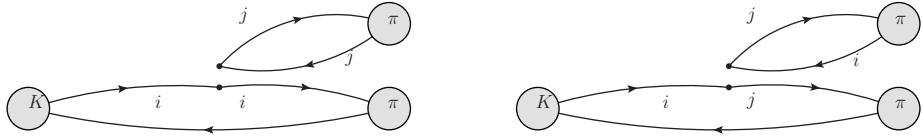}
		\caption{Dominant contributions to Re($A_2$). Re($A_2$) is approximately proportional to the sum of these two contractions. The four-fermion operators have the `left-left' structure and i and j are colour indices.}
		\label{fig:kppi2}
		\end{center}
	\end{figure}

	\begin{figure}
		\begin{center}
		\includegraphics[scale=0.5]{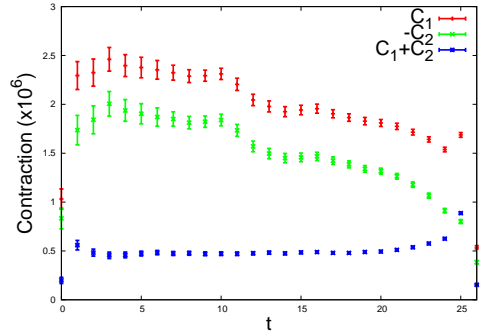}
		\caption{Cancellation of dominant contributions to Re($A_2$) on a $48^3$ lattice with K-$\pi\pi$ separation of 26.}
		\label{fig:c1c2}
		\end{center}
	\end{figure}
	The dominant contribution to Re($A_2$) is the (27,1) operator, which is proportional to the sum of contractions $C_1$ and $C_2$ (see figure \ref{fig:kppi2}).
	One might guess that, if we neglect strong interactions between the pions, that these contractions will be related by $C_1=3C_2$ because of colour suppresion in $C_2$.
	The numerical results for each of these contractions are shown in figure \ref{fig:c1c2}.
	We can see that not only the $C_1=3C_2$ doesn't hold, but that these contractions have opposite signs, which results in a large cancellation in Re($A_2$).
	On the other hand, the contributions to Re($A_0$) which were calculated at threshold (i.e. unphysical kinematics) in \cite{Boyle:2012ys} all have the same sign resulting in an enhancement of Re($A_0$).
	This is a strong hint suggesting an explanation of the $\Delta I = 1/2$ rule, but results from $\Delta I = 1/2$ $K\rightarrow \pi\pi$ calculation are needed to reach a final answer. 

\section{Error budget}
\label{sec:err}
We are currently refining our analysis in order to obtain final results. In the meantime we present an initial estimate for the systematic errors calculated using methods described in~\cite{Blum:2012uk} as well as a comparison with results shown there.
Errors due to lattice artefacts, previously estimated to be about 15\%, are smaller than 4\%.
Since the volume is larger, the exponential errors due to finite volume effects have decreased from 6\% to 2\%.
NPR errors are dominated by perturbative matching to $\overline{\mathrm{MS}}$ scheme and are about 3\% for Re($A_2$) (out of which 0.25\% is the statistical error on renormalization constant) and 6\% for Im($A_2$) (1\% statistical). These are slightly larger than the previous values of 1.8\% for Re($A_2$) and 5.6\% for Im($A_2$).
Errors due to derivative of the phase shift and Wilson coefficients are unchanged and estimated to be approximately 1\% and 6.6\% respecively. 
Adding all estimates of systematic errors in quadrature gives a total systematic error, which is about 11\%. 
This is a significant improvement over the previously quoted 18-19\%.
Contributions from other sources of systematic errors described in \cite{Blum:2012uk} are expected to be negligible.
The dominant contribution the uncertainty comes from the Wilson coefficients, which is a perturbative error.

\section{Conclusions}

We now have the preliminary results for the $A_2$ scattering amplitude at two Iwasaki ensembles, which allows us for the first time to perform the continuum extrapolation for this quantity.
As a consequence, the discretization errors were found to be of order 4\%, which is a significant improvement over the previous estimate of 15\%.

The cancellation between the dominant contractions contributing to Re($A_2$) has been confirmed on both ensembles, which is a significant factor in explaining the $\Delta I = 1/2$ rule.

\acknowledgments{The authors gratefully acknowledge computing time granted through the STFC funded DiRAC facility (grants ST/K005790/1, ST/K005804/1,
ST/K000411/1, ST/H008845/1) as well as the BG/Q computers at the Argonne
Leadership Computing Facility (DOE contract DE-AC02-06CH11357), RIKEN BNL Research Center and Brookhaven National Laboratory.\\
PAB acknowledges support from STFC Grant ST/J000329/1.\\ NHC, RDM, HY and 
DZ were supported in part by US DOE grant 
DE-FG02-92ER40699. }

\end{document}